\documentclass{nuws}
\usepackage{times}
\usepackage{graphicx} 
\begin{document}

\title{Possibility of parametrization of
       atmospheric muon angular flux using underwater data}
\author[1]{S. Klimushin}
\author[1]{E. Bugaev}
\affil[1]{
Institute for Nuclear Research, Russian Academy of Science, Moscow 117312, Russia}
\author[1,2]{I. Sokalski}
\affil[2]{DAPNIA/SPP, CEA/Saclay, 91191 Gif-sur-Yvette CEDEX, France}
\correspondence{klim@pcbai11.inr.ruhep.ru}

\firstpage{1}
\pubyear{2001}

\maketitle

\begin{abstract}
We present the formula for angular distribution of integral flux 
of conventional ($\pi,\,K$) muons deep under water taking into
account the sphericity of the atmosphere and fluctuations of muon energy losses.
The accuracy of this formula for various sea level muon spectra is discussed.
The possibility of reconstructing two parameters of sea level spectrum
by fitting measured underwater angular intensity is shown for Baikal Neutrino
Telescope NT--36 experimental data.
\end{abstract}

\section{Introduction}

The knowledge of expected angular distribution of integral flux of atmospheric muons deep underwater  
is of interest not only for cosmic ray physics 
but also for the estimation of the possible background for neutrino detection and at last for a test 
of the correctness of underwater telescope data interpretation using 
the natural flux of atmospheric muons as calibration source.    
The last item frequently implies the estimation with an appropriate accuracy (e.g., better than 5$\,\%$
for a given sea level spectrum) the 
underwater integral muon flux for various sets of depths, cutoff energies and angular bins 
especially for telescopes of big spacial dimensions.

Up to now the presentation of the results of calculations of muon propagation through thick layers of water 
both for parent muon sea level spectra (especially for angular
dependence taking into account the sphericity of atmosphere) and for underwater angular flux has not been 
quite convenient when applied to concrete underwater arrays.
In addition, a part of numerical results is available only in data tables (often insufficient for accurate interpolation) 
and figures.
The possibility of direct implementation of Monte Carlo methods depends on the availability of corresponding codes
and usually assumes rather long computations and accurate choice of the grid for simulation parameters to avoid big 
systematic errors.   
Therefore, there remains the necessity of analytical expressions for underwater muon integral flux. 
In addition, the possibility of reconstructing the parameters of a sea level
spectrum by fitting measured underwater flux in the case of their direct relation looks rather attractive.

In this paper we present rather simple method allowing one to analytically calculate 
the angular distribution of integral muon flux
deep under water for cutoff energies (1--$10^4)\,$GeV and slant depths of (1--16)$\,$km 
for conventional ($\pi,\,K$) sea level atmospheric muon spectra fitted by means 
of five parameters.  The fluctuations of muon energy losses are taken into account. 

The possibility of reconstructing two parameters of sea level spectrum
by fitting measured underwater angular intensity is shown for Baikal Neutrino
Telescope NT--36 experimental data.

\section{Basic formulas}
According to the approach of work~\citep{KBS} the analytical expression
for calculations of underwater angular integral flux above cutoff energy $E_f$ for a slant depth $R=h/ \cos \theta$ seen at vertical depth $h$ 
at zenith angle $\theta$ and allowing for the fluctuations of energy loss is based on the relation
\begin{equation}\label{ad1}
F_{fl}(\geq E_f,R,\theta)=\frac {F_{cl}(\geq E_f,R,\theta)} {C_{f}(\geq E_f,R,\theta)},  
\end{equation}
where correction factor $C_f$ is expressed, by definition, by the ratio of theoretical integral flux calculated in
the continuous loss approximation to that calculated by exact Monte Carlo, and $F_{cl}(\geq E_f,R,\theta)$
is the angular flux based on continuous energy losses.
 
In principle, the correction factor $C_f$ can be calculated using known 
codes for muon propagation through matter. 
In this work we apply for this aim the MUM code described in work~\citep{MUM}. 

\begin{figure}[t]
\vspace*{2.0mm}
\includegraphics[width=8.3cm]{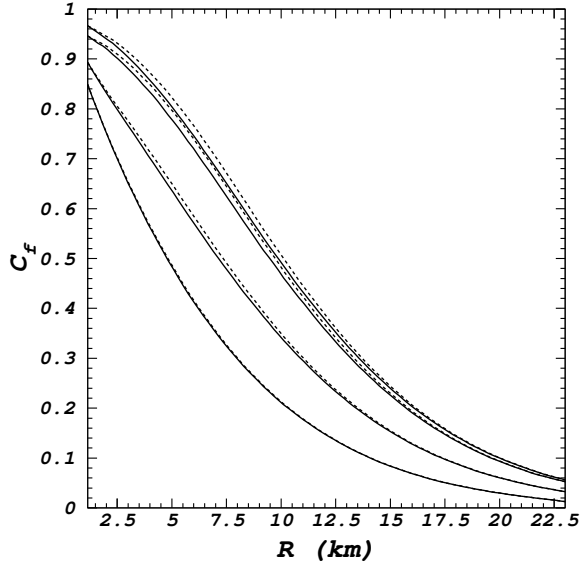} 
\protect\caption{ 
Correction factor $C_f$ as a function of slant depth $R$ in pure water. 
The results obtained using sea level spectrum defined by expression~(6) are given.
Solid curves correspond to numerical calculations for vertical case $\theta=0^\circ$. 
Dashed curves describe the correction factor computed at vertical depth $h$ of 1.15 km for various zenith angles
as a function of slant depth defined by $R=h/\cos\theta$. 
Both solid and dashed curves are shown for four values of cut-off energy 
$E_f$: 10 GeV, 100 GeV, 1 TeV, and 10 TeV, from top to bottom.
\label{fig:cf_bk}}
\end{figure}

\begin{table*}
\protect\caption{ Coefficients $c_{ij}$ of the fitting formula~(\protect\ref{CF}) for correction factor calculated for vertical
                  sea level spectrum given by expression~(6) below.    
\label{tab:cftab}}
\center{\begin{tabular}{crrrrr} \hline \hline
subscript $i$  & $c_{i0}$~~~~~~~ & $c_{i1}$~~~~~~~ & $c_{i2}$~~~~~~~ & $c_{i3}$~~~~~~~ & $c_{i4}$~~~~~~~ \\\hline
0              &  $ 6.3045 \times10^{-1}$ & $ 6.6658 \times10^{-1}$ & $-4.5138 \times10^{-1}$ & $ 1.2441 \times10^{-1}$ & $-1.1904 \times10^{-2}$ \\ 
1              &  $ 2.0152 \times10^{-1}$ & $-4.2990 \times10^{-1}$ & $ 3.2532 \times10^{-1}$ & $-1.0265 \times10^{-1}$ & $ 1.0751 \times10^{-2}$ \\ 
2              &  $-3.3419 \times10^{-2}$ & $ 5.1833 \times10^{-2}$ & $-3.9229 \times10^{-2}$ & $ 1.2360 \times10^{-2}$ & $-1.2911 \times10^{-3}$ \\ 
3              &  $ 1.6365 \times10^{-3}$ & $-2.3645 \times10^{-3}$ & $ 1.7775 \times10^{-3}$ & $-5.5495 \times10^{-4}$ & $ 5.7557 \times10^{-5}$ \\ 
4              &  $-2.6630 \times10^{-5}$ & $ 3.7770 \times10^{-5}$ & $-2.8207 \times10^{-5}$ & $ 8.7275 \times10^{-6}$ & $-8.9919 \times10^{-7}$ \\ 
\hline \hline
\end{tabular}}
\end{table*} 
The values of correction
factors calculated for the same slant depth $R$ at vertical direction and at zenith angle $\theta$ differ weakly. 
It is illustrated in Fig.~\ref{fig:cf_bk}, where one can see that $C_{f}(\geq E_f,R,0^{\circ})$ differs from  
$C_{f}(\geq E_f,R,\arccos h/R)$ maximum on 3.3$\,\%$ for $E_f>$10 GeV at vertical depth $h$ of 1.15 km. 
It appears that with acceptable accuracy the correction factor depends on slant depth $R$ only, rather than on $R$ and $\theta$
separately. 

The dependencies of correction factor on $E_f$ and $R$, calculated for sea level spectrum
given by expression~(6) below represent the set of rather smooth curves (shown in Fig.~\ref{fig:cf_bk}) 
and it is possible to approximate this factor by formula  
\begin{equation}\label{CF}
C_f(\geq E_f,R,\theta)=\sum_{i=0}^4 ( \sum_{j=0}^4 c_{ij}\log^{j}_{10}\,E_{f} )R^i. 
\end{equation}
Here cut-off energy $E_f$ is expressed in (GeV) and slant depth $R$ is in (km) with the coefficients $c_{ij}$
collected in Table~\ref{tab:cftab}. When using~(\ref{CF}) for cutoff energies $E_f<$10 GeV one should substitute value of $E_f$=10 GeV.

Formula~(\ref{CF}) can be applied for any geometrical shape of the surface. Right hand side of~(\ref{CF}) depends on
$\theta$ because, generally, $R=R(\theta)$. So, in the particular case of a flat surface the angular dependence of the
correction factor appears, in our approximation, only through the relation  
$R=h / \cos\theta$
~(where $h$ is a vertical depth). 

The accuracy of formula~(\ref{CF}) for $E_f$=(1--100)$\,$GeV is better than $\pm2\,\%$ for slant depths $R$ as large as 22 km and
is not worse than $\pm3\,\%$ for $E_f$=1 TeV up to $R$=17 km and for $E_f$=10 TeV up to $R$=15 km.
Fig.~\ref{fig:cf_bk} shows that for $E_f<\,$100 GeV the total energy loss may be treated as quasi-continuous  
(at level of $C_f>\,$0.9) only for slant depths $R<\,$2.5 km but for $E_f$=10 TeV the fluctuations should be taken into account
at level of 15$\,\%$ already for slant depth as small as $R$=1 km. 

\begin{figure}[t]
\vspace*{2.0mm} 
\includegraphics[width=8.3cm]{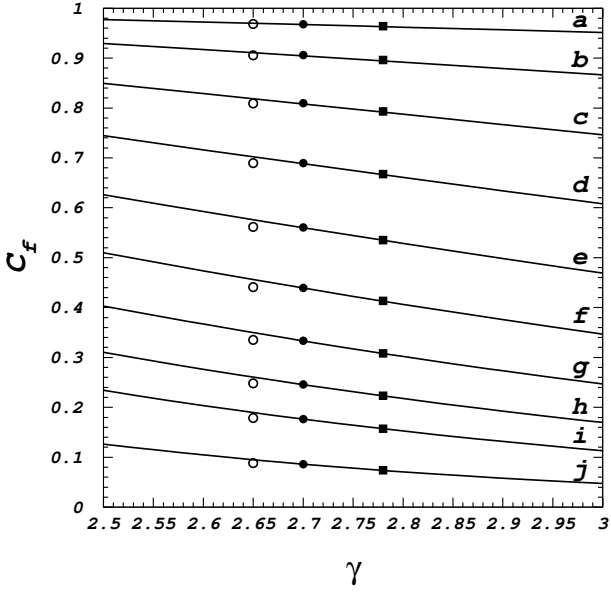} 
\protect\caption{ 
Correction factor $C_f$ as a function of spectral index $\gamma$ of sea level spectrum for various depths in water for vertical direction.
The distributions for cut-off energy $E_f$=10 GeV are given.
Solid curves correspond to numerical computations using sea level spectrum defined by Eq.~(\protect\ref{BKfit}) with varying 
spectral index $\gamma$. Open circles correspond to numerical computations using VZK sea level spectrum~\citep{VZK},
closed circles -- Gaisser's sea level spectrum~\citep{Gaisser}, squares -- MACRO~\citep{MACRO} sea level spectrum.
All distibutions are shown for the following values of vertical depth in pure water: 1.15 km (a), 3 km (b), 5 km (c), 7 km (d),
9 km (e), 11 km (f), 13 km (g), 15 km (h), 17 km (i), and 21 km (j), from top to bottom.
\label{fig:cf_gamma}}
\end{figure}

The dependence of correcton factor $C_f$ on different sea-level vertical spectra 
is illustrated by Fig.~\ref{fig:cf_gamma}. The correction factors calculated for $E_f$=10 GeV using sea level spectrum   
~(\ref{BKfit}) with spectral index $\gamma$ of 2.5 and 3.0 (instead of 2.72) differ more than on a factor of 2 starting from 
slant depth of $R$=12 km. Nevertheless, the values of $C_f$ calculated using sea level spectra having $\gamma$=(2.65--2.78) 
are already within $\pm5\,\%$ corridor. For $E_f$=1 TeV this corridor is larger on $2\,\%$.
This fact results in the possibility to extrapolate the parametrization~(\ref{CF}) based on sea level spectrum having $\gamma$=2.72 
to other spectra at least up to slant depths of (12--13) km without introduction 
of additional spectral corrections.  

The angular flux $F_{cl}(\geq E_f,R,\theta)$ based on effective 
linear continuous energy losses $\alpha + \beta E$ having 2 slopes, is calculated by the following rule: 
\begin{eqnarray} \label{ad2}
F_{cl}(\geq E_f,R,\theta) = \nonumber \\ 
\left\{
\begin{array}{ll} F_{cl}(\geq E_f,R,\theta;\alpha_{1},\beta_{1})
                         & \mbox{~for~}  R \leq R_{12}, \\
                F_{cl}(\geq E_{12},(R-R_{12}),\theta;~\alpha_{2},\beta_{2})
                         & \mbox{~for~}  R > R_{12}. 
\end{array}\right.
\end{eqnarray}
Here $E_{12}$ is the energy in the point of slope change from $(\alpha_1,\beta_1)$ to $(\alpha_2,\beta_2)$ and 
$R_{12}$ is the muon path from the energy $E_{12}$ till $E_f$ which is given by
$$R_{12}=\frac {1}{\beta_1}\,\ln \biggl( \frac {\alpha_{1}+E_{12}\beta_{1}}{\alpha_{1}+E_{f}\beta_1} \biggr).$$ 

The formula for integral muon angular flux in the assumption of linear 
continuous energy losses is as follows:
\begin{eqnarray}\label{ad3}
F_{cl}(\geq E_f,R,\theta;~\alpha,\beta)
=\frac{e^{-\beta R\gamma}}{\gamma} \nonumber \\ 
\times \sum_{i={\pi,K}} D_{0_i} E^{cr}_{0_i}(\theta) 
(E_f+y_i)^{-\gamma} (1-z_i)^{1-\gamma}\,\mbox{S}(z_{i},\gamma), 
\end{eqnarray}
where subscript $i$ stands over both pion ($\pi$) and kaon ($K$) terms and 
\begin{eqnarray*}
y_i = \frac{\alpha}{\beta}\,(1-e^{-\beta R}) + E^{cr}_{0_i}(\theta)\,e^{-\beta R}, \\
z_i = \frac {E^{cr}_{0_i}(\theta)\,e^{-\beta R}}{ E_{f}+y_i },
\qquad E^{cr}_{0_i}(\theta)=\frac {E^{cr}_{0_i}(0^\circ)}{\cos\theta^*},\\
\mbox{S}(z,\gamma) = 1+ \sum_{n=1}^{\infty} n!\,z^{n} \biggl (\prod_{j=1}^{n} (\gamma+j) \biggr )^{-1}= 
1+\frac{z}{\gamma+1}\\
+\frac{2z^2}{(\gamma+1)(\gamma+2)}+ 
\frac{6z^3}{(\gamma+1)(\gamma+2)(\gamma+3)} + \dots~. 
\end{eqnarray*}

When using expression~(\ref{ad3}) for slant depths $R>R_{12}$ one must substitute $R \to (R-R_{12})$ 
and $E_f \to E_{12}$ and use the values ($\alpha_2,\beta_2$) for a loss description. For slant depths $R \leq R_{12}$ the use of~
(\ref{ad3}) remains unchangeable and the loss values are expressed by ($\alpha_1,\beta_1$).
This algorithm may be extended to computations with any number of slopes of the energy losses.

The 5 parameters ($D_{0_\pi},D_{0_K},E^{cr}_{0_\pi}(0^\circ),E^{cr}_{0_K}(0^\circ),\gamma$) are those of the 
differential sea level muon spectrum, for which we use the following parametrization:
\begin{equation}\label{Sealds1}
D(E_{0},\theta)= E_{0}^{-\gamma} \sum_{i={\pi,K}} \frac{D_{0_i}}{1+E_{0}/E^{cr}_{0_i}(\theta)},  
\end{equation}
where $\gamma$ is a spectral index and $E^{cr}_{0_{\pi,K}}(\theta)$ have approximate sense of critical energies 
of pions and kaons for given zenith angle and $E^{cr}_{0_{\pi,K}}(0^\circ)$ are those for vertical direction.
The corresponding angular distrubution should be introduced using an analytical description of effective cosine $\cos\theta^*$  
taking into account the sphericity of atmosphere. It should be noted that the description of underwater angular flux with the
5 parameters of a sea level spectrum gives the possibility of their direct best fit using the 
experimental underwater distribution.

Flux value in~(\ref{ad3}) is expressed in units of (cm$^{-2}$s$^{-1}$sr$^{-1}$) and all energies are in (GeV),
slant depth $R$ in units of (g$\,$cm$^{-2}$), loss terms $\alpha$ and $\beta$ 
in units of $(10^{-3}$GeVcm${}^2$g${}^{-1})$ and $(10^{-6}\mbox{cm}^2\mbox{g}^{-1})$, correspondingly.
For the description of effective linear continuous energy losses we use the following values of parameters
when substituting in~(\ref{ad2}): ($\alpha_1$=2.67, $\beta_1$=3.40) and ($\alpha_2$=$-$6.5, $\beta_2$=3.66) with $E_{12}$=35.3 TeV.

To examine the angular behaviour of a flux given by the formula~(\ref{ad1}) by means of the comparison with numerical calculations 
we used the following parameters of the sea level muon spectrum: 
\begin{eqnarray*}
D_{0_\pi}=0.175,\qquad  D_{0_K}=6.475 \times 10^{-3},\\
E^{cr}_{0_{\pi}}(0^\circ)=103~\mbox{GeV},\quad E^{cr}_{0_{K}}(0^\circ)=810~\mbox{GeV}, \quad \gamma=2.72~.
\end{eqnarray*}
These values have been chosen according to splines computed in this work via the data tables kindly given us by 
authors of Ref.~\citep{Sineg}. 
When checking the values of fit spectrum for $\cos\theta$=(0.05--1.0) 
we realized that the standard description of effective cosine (with geometry of spherical
atmosphere and with definite value of effective height of muon generation) is not enough
and one should introduce an additional correction $S(\theta)$ leading to (10--20)$\,\%$
increase of effective cosine value for $\cos\theta<$~0.1. The reason of an appearing of this correction is that the concept of an
effective generation height is approximate one. It fails at large zenith angles where the real geometrical size of the generation region
becomes very large.

\begin{figure}[!t]
\includegraphics[width=8.3cm]{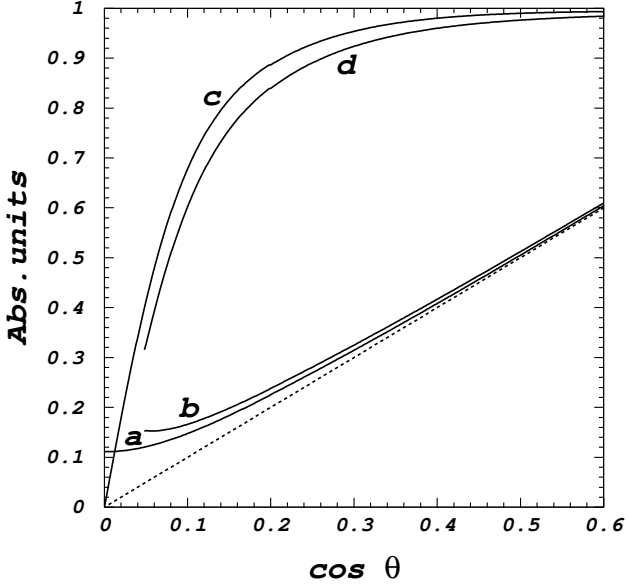}
\protect\caption{
Effective cosine as a function of zenith angle.
Curve (a) is geometrical effective cosine $\cos\theta^*$ given 
by Eq.~(\protect\ref{Efcosfit}).
Curve (b) is effective cosine $\cos\theta^{**}$ with the correction and is given by Eq.~(\protect\ref{Efcos}).
Curves (c) and (d) represent the ratio $\cos\theta/\cos\theta^*$ and $\cos\theta/\cos\theta^{**}$, correspondingly. 
\label{fig:efcos}}
\end{figure}

The resulting fit of angular sea level spectrum in units of~(cm$^{-2}$s$^{-1}$sr$^{-1}$Ge$V^{-1}$) is given by
\begin{eqnarray}\label{BKfit}
D(E_0,\theta)=0.175 E_0^{-2.72} \nonumber \\
\times \left(\frac{1}{\displaystyle 1+\frac{E_0\cos\theta^{**}}{\displaystyle 103}}
+\frac{0.037}{\displaystyle 1+\frac{E_0\cos\theta^{**}}
{\displaystyle 810}}\right),
\end{eqnarray}
with modified effective cosine expressed by
\begin{equation}\label{Efcos}
\cos\theta^{**} =S(\theta)\cos\theta^{*},  
\end{equation}
where $\cos\theta^*$ is derived from spherical atmosphere geometry and is given by 
the polynomial fit:  
\begin{equation}\label{Efcosfit}
\cos\theta^{*}=\sum_{i=0}^4 c_{i}\cos^{i}\theta,  
\end{equation}
with the coefficients of the decomposition assembled in Table~\ref{tab:ectab}.
The accuracy of~(\ref{Efcosfit}) is much better than 0.3$\,\%$ except the region $\cos\theta$=(0.3--0.38) where
it may reach the value of 0.7$\,\%$.  
Note that for $\cos\theta>$~0.4 the influence of the curvature of real atmosphere is less
than 4~$\%$ but for $\cos\theta<$~0.1 it is greater than 40~$\%$ (Fig.~\ref{fig:efcos}). 

$S(\theta)$ is the correction which is given for $\sec\theta\leq20$ by
\begin{equation}\label{Scos}
S(\theta)=0.986+0.014\sec\theta.  
\end{equation}

Correspondingly, for critical energies in expression (6) 
one should use $\cos\theta^{**}$ instead of $\cos\theta^{*}$. 


The energy region, inside which the deviation of angular spectrum given by Eq.~(\ref{BKfit}) from 
parent one is less than 5~$\%$, is shifted from    
(0.3--200)$\,$TeV for $\cos\theta$=1.0 to (1.5--300)$\,$TeV for $\cos\theta$=0.05.  
The sea level spectrum given by~(\ref{BKfit}) is valid only below the knee ($E_{0}\sim\,$300 TeV) of primary cosmic ray spectrum.

\begin{table*}[htb]
\protect\caption{Coefficients $c_{i}$ of the fitting formula~(\protect\ref{Efcosfit})
                 for effective cosine with the maximum relative errors. }
\label{tab:ectab}
\center{\begin{tabular}{ccccccc} \hline \hline
$\cos \theta$ & $c_0$  & $c_1$ & $c_2$ & $c_3$ & $c_4$ & Max.err,$\%$ \\\hline
0$\div$0.002    & 0.11137 & 0 & 0 & 0 & 0 & 0.004\\ 
0.002$\div$0.2  & 0.11148 & $-0.03427$ & 5.2053  & $-14.197$ & 16.138 & 0.3\\ 
0.2$\div$0.8    & 0.06714 & 0.71578  & 0.42377 & $-0.19634$ & $-0.021145$ & 0.7\\\hline \hline 
\end{tabular}}
\end{table*} 
\section{Comparison with numerical calculations}
\begin{figure}[t]
\vspace*{2.0mm} 
\includegraphics[width=8.3cm]{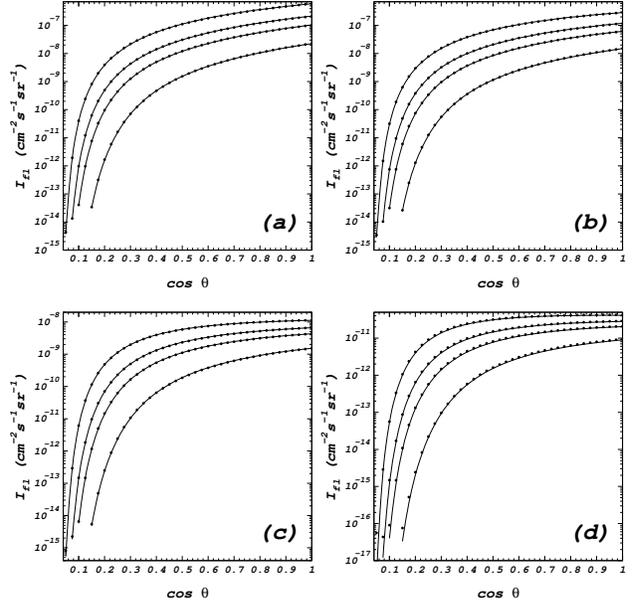}
\protect\caption{ 
Underwater integral muon flux allowing for loss fluctuations as a function of zenith angle at different vertical depths.
Four pictures are shown for various cut-off energies $E_f$: 10 GeV (a), 100 GeV (b), 1 TeV (c), and 10 TeV (d), correspondingly. 
Four curves at each picture correspond to vertical depths $h$: 1.15 km, 1.61 km, 2.0 km, and 3.0 km, from top to bottom. 
Solid curves result from numerical computations using  
the sea level spectrum based on data tables from~\citep{Sineg} 
 and MUM code of muon propagation. Dotted curves result from analytical
expression~(\protect\ref{ad1}) using the sea level spectrum~(\protect\ref{BKfit}). 
\label{fig:Ifl_all}}
\end{figure}
\begin{figure}[t] 
\includegraphics[width=8.3cm]{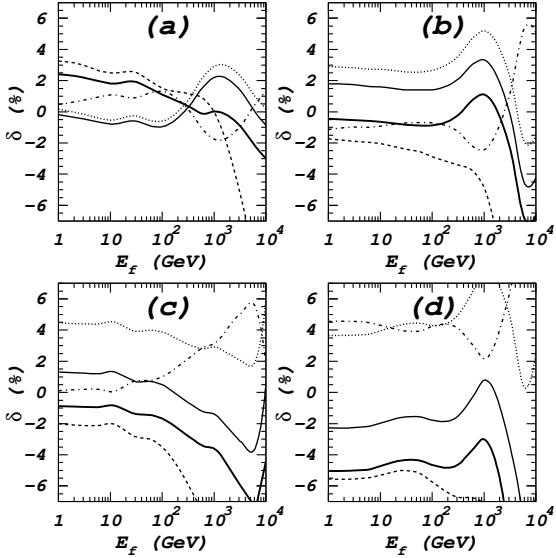}

\protect\caption{
The accuracy of formula~(1) as function of cutoff energy $E_f$ for different 
vertical sea level spectra. Four pictures are shown for various 
vertical depths $h$: 2 km (a), 5 km (b), 10 km (c), and 15 km (d), correspondingly. 
Thick solid curves correspond to comparison with numerical computations using sea level muon spectrum 
based on data tables from~\citep{Sineg} and MUM code of muon propagation. Thin solid curves --  
sea level spectrum defined by Eq.~(\protect\ref{BKfit}), dashed -- VZK sea level spectrum~\citep{VZK},
dash-dotted -- Gaisser's sea level spectrum~\citep{Gaisser}, dotted -- LVD~\citep{LVD} sea level spectrum.
\label{fig:errors}}
\end{figure}
The examination of~(\ref{ad3}) showed rather quick convergence of series S$(z,\gamma)$ with increase of $R$ and $E_f$. 
Therefore, for the accuracy of $F_{cl}$ computation better than 0.1~$\%$ it is quite enough to take only 
four first terms of this series (up to $z^3$) for all values $R>$~1 km and $E_f$ in (1--10$^{4}$) GeV. Even using the two
terms leads to the accuracy of 1.3~$\%$ for ($R$=1.15 km, $E_f$=1 GeV) and $<$0.5$\,\%$ for ($R>\,$2.5 km, $E_f>\,$1 GeV).

Fig.~\ref{fig:Ifl_all} shows the comparison of underwater angular integral fluxes allowing for loss fluctuations 
at different basic depths $h$ (of location of existing and planned telescopes) calculated both numerically 
using MUM code~\citep{MUM} for parent sea level spectrum  
and analytically~(\ref{ad1}) for the spectrum given by~(\ref{BKfit}). 

We realized that the error given by formula~(\ref{ad1}) for     
all mentioned sea level spectra is within the corridor of $\pm$(4--6)$\,\%$ for all cutoff 
energies $E_f$=(1--10$^{3})\,$GeV and
slant depths $R$=(1--16)$\,$km (corresponding angle is expressed by $\cos\theta=h/R$ for a given vertical depth $h$). This is proved
for $h$ in a range (1--3)$\,$km. 
For bigger cutoffs of $E_f$=(1--10)$\,$TeV the corridor of errors is $\pm$(5--7)$\,\%$ for $R$=(1--13)$\,$km. Note that for the
sea level spectrum~(\ref{BKfit}), just used for $C_f$ parametrization, the errors are smaller on 2$\,\%$. 

The accuracy of the parametrization, used for the correction factor as a function of $E_f$ and slant depth $R$   
is rather high and is about $\pm5\,\%$ for all angles and kinds of the sea level spectrum (assuming that the spectral
index $\gamma$ is approximately within (2.65--2.78)) (Fig.~\ref{fig:errors}). 
It results in the possibility to use it for an estimating
numerically from various sea level spectra the value of an angular integral flux allowing for fluctuations of losses without
direct Monte Carlo simulations. 

Note that the expression~(\ref{ad1}) may be directly used for an ice after the substitution $R \to R/ \rho$, with $\rho$ 
being the ice density, and, with an additional error of $\sim2\,\%$, for sea water. In spite of seeming complexity of the formulas~(\ref{ad1}),
~(\ref{ad2}) and~(\ref{ad3}) they may be easily programmed.

The validity of this analytical expression with an accuracy of $\pm$(5--7)$\,\%$ for $E_f$=(10$^3$--10$^4$)$\,$GeV 
and slant depths of (1--12)$\,$km gives also the possibility of estimation the angular underwater differential 
spectrum (by means of numerical differentiation) with error smaller than $\pm$(6--8)$\,\%$ for 
energies of (30--5$\times10^{3}$)$\,$GeV.

\section{Parametrization of atmospheric muon angular flux using underwater data}
\label{sec:NT-36}
When reconstructing the parameters of sea level spectrum defined by Eq.~(5)  
by fitting with MINUIT least square method the corresponding underwater angular intensity 
expected at vertical depth $h$=1.15$\,$km and expressed by formula described in Sec.~1  we have realized that:
\begin{itemize}
\item[(i)] it is possible to reconstruct two parameters ($D_{0_\pi},\gamma$) of sea level spectrum
when angular bins corresponding to slant depth $R \geq$6$\,$km are involved
\item[(ii)] the reconstruction of third parameter $D_{0_K}$ is formally possible only using
angular bins corresponding to slant depths $R \geq$15$\,$km where neutrino induced intensity
should be taken into account.
\end{itemize}

\begin{figure}[t]
\vspace*{2.0mm} 
\includegraphics[width=8.3cm]{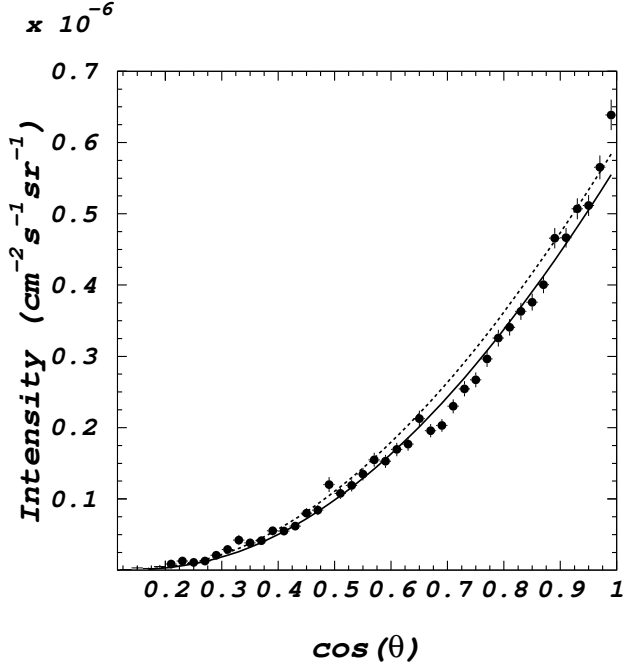}
\protect\caption{ 
Zenith angle distribution of the muon intensity at vertical depth of 1.15$\,$km.
Experimental points - NT-36 data~\citep{NT-200}. Solid curve results from the best fit of 2 parameters 
($D_{0_\pi}=0.26,\gamma=2.79$) of sea level spectrum. Dashed curve results 
from analytical expression~(\protect\ref{ad1}) using the sea level spectrum~(\protect\ref{BKfit})
and is consistent with experimental data also with CL=0. 
Only statistical errors have been taken into account.
\label{fig:nt36_1}}
\end{figure}

\begin{figure}[t]
\vspace*{2.0mm} 
\includegraphics[width=8.3cm]{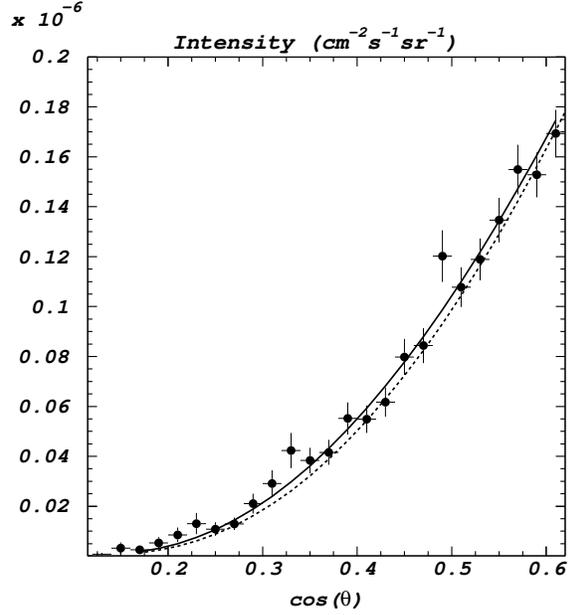}
\protect\caption{ 
Zenith angle distribution of the muon intensity at vertical depth of 1.15$\,$km for horizontal directions.
Experimental points - NT-36 data~\citep{NT-200}. Solid curve results from the best fit of 2 parameters 
($D_{0_\pi}=0.12,\gamma=2.68$) of sea level spectrum. Dashed curve results from the best fit of 2 parameters 
($D_{0_\pi}=0.26,\gamma=2.79$) of sea level spectrum using a whole angular range. 
Only statistical errors have been taken into account.
\label{fig:nt36_2}}
\end{figure}

For checking the same procedure using experimental results we have examined 
the data sample with NT-36 (1993) unfolded experimental angular intensity 
published by Baikal Collaboration in Ref.~\citep{NT-200} for vertical depth of $h$=1.15$\,$km. 
The cutoff energy value was taken as $E_f$=10$\,$GeV.  
The whole data sample corresponds to 44 angular bins $\Delta\cos\theta$=0.02 ($\cos\theta$=(0.13--0.99))
with maximum slant depth $R$=8.8$\,$km. The mean muon energy at the sea level corresponding to
this angular range is $E$=(0.6--15)$\,$TeV. 
Only statistical errors have been taken into account.
The following 3 parameters of sea level spectrum were taken according to expressions~(5) and~(\ref{BKfit}): 
\begin{eqnarray*}
D_{0_K}=0.037 D_{0_\pi},\\
E^{cr}_{0_{\pi}}(0^\circ)=103~\mbox{GeV},\quad E^{cr}_{0_{K}}(0^\circ)=810~\mbox{GeV}.
\end{eqnarray*}
The results of reconstructing of two free parameters ($D_{0_\pi},\gamma$) of sea level spectrum are as follows.
\begin{itemize}
\item[(i)] For a range of zenith angles within $\cos\theta$=(0.17--0.99) we have obtained 
formally ($D_{0_\pi}=0.26,\gamma=2.79$). It is illustrated by Fig.~\ref{fig:nt36_1}. 
In spite of this result coincides with MACRO~\citep{MACRO} and LVD~\citep{LVD}
best fits, its confidence level (CL) is close to 0. The artificial increase of errors in 3 times
due to additional systematic errors leads to ($D_{0_\pi}=0.17,\gamma=2.73$) with CL=87$\,$\%. 
\item[(ii)] For vertical directions with $\cos\theta$=(0.61--0.99) the reconstructed sea level spectrum
is extremely steep with ($D_{0_\pi}=1.0,\gamma=3.0$) and CL=0.5$\,$\% but
the increase of errors in 2 times results in ($D_{0_\pi}=0.19,\gamma=2.74$) with CL=40$\,$\%. 
\item[(iii)] For horizontal directions with $\cos\theta$=(0.13--0.61) the reconstructed sea level spectrum
is flat, as ($D_{0_\pi}=0.1,\gamma=2.65$) with CL=70$\,$\%, and for $\cos\theta$=(0.17--0.61)
as ($D_{0_\pi}=0.12,\gamma=2.68$) with CL=40$\,$\%. The result of this best fit is shown in
Fig.~\ref{fig:nt36_2}. 
\end{itemize}
It should be pointed out that the implementation of Gaisser's set of 3 
parameters ($D_{0_K},E^{cr}_{0_\pi}(0^\circ),E^{cr}_{0_K}(0^\circ)$)~\citep{Gaisser} gives
almost the same results of reconstructing of ($D_{0_\pi},\gamma$), as well as when using 
the recalculated depth-intensity curve. 
The fact that sea level spectrum changes the slope from vertical directions to horizontal
ones may be explained probably by unproper taking into account the muon bundles when unfolding
the measured intensity.  

\section{Conclusions}
\label{sec:Concl}
The analytical expression presented in this work allows to estimate for fluctuating losses the integral flux of
atmospheric muons in pure water expected for different zenith angles, $\cos\theta$=(0.05--1.0), at various vertical
depths at least of $h$=(1--3)$\,$km for different parametrizations of the sea level muon spectra. 
The errors of this expression are estimated to
be smaller than $\pm$(4--6)$\,\%$ for cutoff energies ranged in $E_f$=(1--10$^{3})\,$GeV and slant depths 
in $h/\cos\theta$=(1--16)$\,$km. The main advantage of the presented 
formula consists in the possibility of the direct best fit of at least 2
parameters of parent sea level spectrum using angular distribution of underwater integral flux measured experimentally at a given
vertical depth.  

The fitted sea level spectrum for NT-36 data is too steep for vertical directions
($\gamma$=3.0) and flat for horizontal ones ($\gamma$=2.65--2.68). It leads to the necessity of
proper introducing of systematic errors mainly resulted from muon bundles.
The artificial increase of statistical errors in 2--3 times results in sea level spectra closer
to ~\citep{KBS} and ~\citep{Gaisser}.

The proposed method may be adapted to estimations in rock after 
corresponding description of the correction factor and continuous effective losses.

\end{document}